\documentclass[superscriptaddress,reprint,amsmath,amssymb,aps,prl]{revtex4-2}
\usepackage[english]{babel}
\usepackage{amsmath,amssymb,mathtools,float}
\usepackage{graphicx}
\usepackage{dcolumn}
\usepackage{bm}
\usepackage{xr}
\usepackage[version=4]{mhchem}
\usepackage[hidelinks]{hyperref}
\usepackage[dvipsnames]{xcolor}
\usepackage{siunitx}
\DeclareSIUnit\angstrom{\text {Å}}
\DeclareSIUnit\molar{\text {M}}
\usepackage{layouts}

\usepackage[capitalize]{cleveref}

\makeatletter

\begin{document}

\title{Water, not salt, causes most of the Seebeck effect of nonisothermal aqueous electrolytes}

\author{Ole Nickel}
\affiliation{Institute of Polymers and Composites, Hamburg University of Technology, Hamburg, Germany}

\author{Ludwig J. V. Ahrens-Iwers}
\affiliation{Institute of Advanced Ceramics, Hamburg University of Technology, Hamburg, Germany}

\author{Robert H. Mei{\ss}ner}
\email{robert.meissner@tuhh.de}
\affiliation{Institute of Polymers and Composites, Hamburg University of Technology, Hamburg, Germany}
\affiliation{Institute of Surface Science, Helmholtz-Zentrum Hereon, Geesthacht, Germany}

\author{Mathijs Janssen}
\email{mathijs.a.janssen@nmbu.no}
\affiliation{Norwegian University of Life Sciences, Faculty of Science and Technology, Pb 5003, 1433, \AA s, Norway}

\date{\today}

\begin{abstract}
When two electrolyte-immersed electrodes have different temperatures, a voltage $\Delta \psi$ can be measured between them.
This electrolyte Seebeck effect is usually explained by cations and anions flowing differently in thermal gradients.
However, our molecular dynamics simulations of aqueous electrolytes reveal a large temperature-dependent potential drop $\chi$ near blocking electrodes caused by water layering and orientation. 
The difference in surface potentials at hot and cold electrodes is more important to the Seebeck effect than ionic thermodiffusion,  $\Delta \psi \sim \chi_{\rm hot}-\chi_{\rm cold}$.
\end{abstract}

\maketitle

Industries discard thermal energy on a large scale, and tapping into this resource may help society with its much-needed energy transition \cite{jouhara_waste_2018}.
Among the alternatives, electric and electrochemical devices with temperature-induced open-circuit voltages are attractive as they have no moving parts \cite{wang_ionic_2017,yang_ionic_2021,zhao2021ionic}. 
The generation of a thermovoltage $\Delta \psi=\psi_\mathrm{hot}-\psi_\mathrm{cold}$ by a device subject to a temperature difference $\Delta T=T_\mathrm{hot}-T_\mathrm{cold}$ is called the Seebeck effect, characterized by the Seebeck coefficient $\mathcal{S} = - \Delta \psi / \Delta T$ \cite{dupont_thermo-electrochemical_2017, yang_ionic_2021}.
The Seebeck effect of solid-state devices relies on electrons and holes moving apart in thermal gradients \cite{zhang_thermoelectric_2015}. 
Electrochemical cells filled with aqueous \cite{sehnem_time-dependent_2021} and polymeric \cite{zhao_ionic_2016} electrolytes and ionic liquid/organic solvent mixtures \cite{bonetti_huge_2011, bonetti_thermoelectric_2015} show a Seebeck effect as well, which is usually explained by anions and cations moving apart in thermal gradients.
Faradaic processes can also cause a Seebeck effect in cells  with redox-active electrolytes \cite{dupont_thermo-electrochemical_2017, abraham_seebeck_2011}.

Ions move in nonisothermal fluids due to the interactions among themselves and with solvent molecules \cite{wiegand_thermal_2004,zhao2021ionic}. 
On mesoscopic length scales, the thermodiffusion flux of an ion species $i$ can be written as $J_i^{\rm th}=- Q_i^\ast \nabla T $, with $Q_i^{\ast}$ being the ions' heat of transport and $T$ being temperature.
The $Q_i^{\ast}$ of aqueous electrolytes relate to the Gibbs free hydration energy $G^\mathrm{hyd}_i$ and hydration entropy $S^\mathrm{hyd}_i$ \cite{rezende_franco_molecular_2021, takeyama_proportionality_1988} by 
\begin{equation}
    \frac {Q_i^\ast} {T} = \frac{\mathrm{d}G^\mathrm{hyd}_i}{\mathrm{d}T} = - S^\mathrm{hyd}_i.
    \label{eq:soret-entropy}
\end{equation}

\begin{figure}
    \centering
    \includegraphics[width=\linewidth]{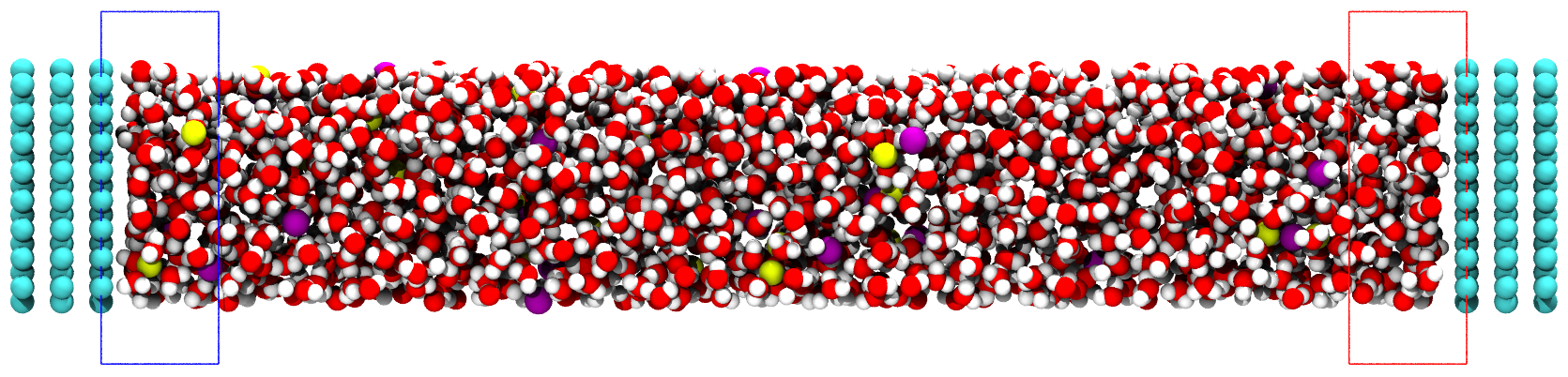}
    \caption{Simulation snapshot of \SI{1}{\molar} LiI in water between graphene electrodes. Thermostats control the temperature in the blue and red regions independently. ($\mathrm{Li^{+}}$ - yellow, $\mathrm{I^{-}}$ - purple, hydrogen - white, oxygen - red, carbon - cyan). } 
    \label{fig:picture_cell}
\end{figure}
First, $S^\mathrm{hyd}_i$ is negative (and $Q_i^{\ast}$ positive) for most ions, so when electrodes block fluxes, cations and anions pile up near the cold electrode.
This so-called Soret effect of spatially varying salt concentration can be probed through its impact on the electrolyte's conductivity \cite{leaist_heats_1994, agar_new_1960, snowdon_concentration_1960} and refractive index \cite{romer_alkali_2013, colombani_hydrodynamic_1998}, giving experimental access to $Q_i^{\ast}$. 
Second, the size and valency of ions affect their hydration shell and, thus, their hydration entropy.
According to Marcus's theory \cite{marcus_simple_1994}, smaller ions have larger hydration shells and more negative $S^\mathrm{hyd}_i$.
Third, $Q_i^{\ast}$ tends to be larger for cations than for anions \cite{agar_single-ion_1989}. 
If so, thermodiffusion of an electrolyte between electrodes at different temperatures leads to excess cations near the cold and anions near the hot electrode.
This ionic charge separation causes a potential drop between the electrodes: the Seebeck effect.
Combining $J_i^{\rm th}$ and the Nernst-Planck equation for ionic fluxes due to diffusion and electromigration, an expression can be derived for the steady-state Seebeck coefficient \cite{eastman_theory_1928,wurger_thermal_2010,romer_alkali_2013,sehnem_time-dependent_2021}. 
For binary electrolytes, 
\begin{equation}
    \mathcal{S} = \frac {Q_+^\ast-Q_-^\ast} {2eT}\,,
    \label{eq:seebeck}
\end{equation}
where $e$ is the elementary charge.
As \cref{eq:seebeck} followed from an extended Nernst-Planck equation, it accounts for ionic thermodiffusion and mean-field electrostatic interactions. 
Finite ion sizes and ion-ion correlations, important for dense electrolytes and ionic liquids \cite{Kornyshev2007, Kondrat2011, lee_ion_2021}, are ignored, so \cref{eq:seebeck} may not hold for such fluids.

\begin{figure*}
    \centering
    \includegraphics[width=1\textwidth]{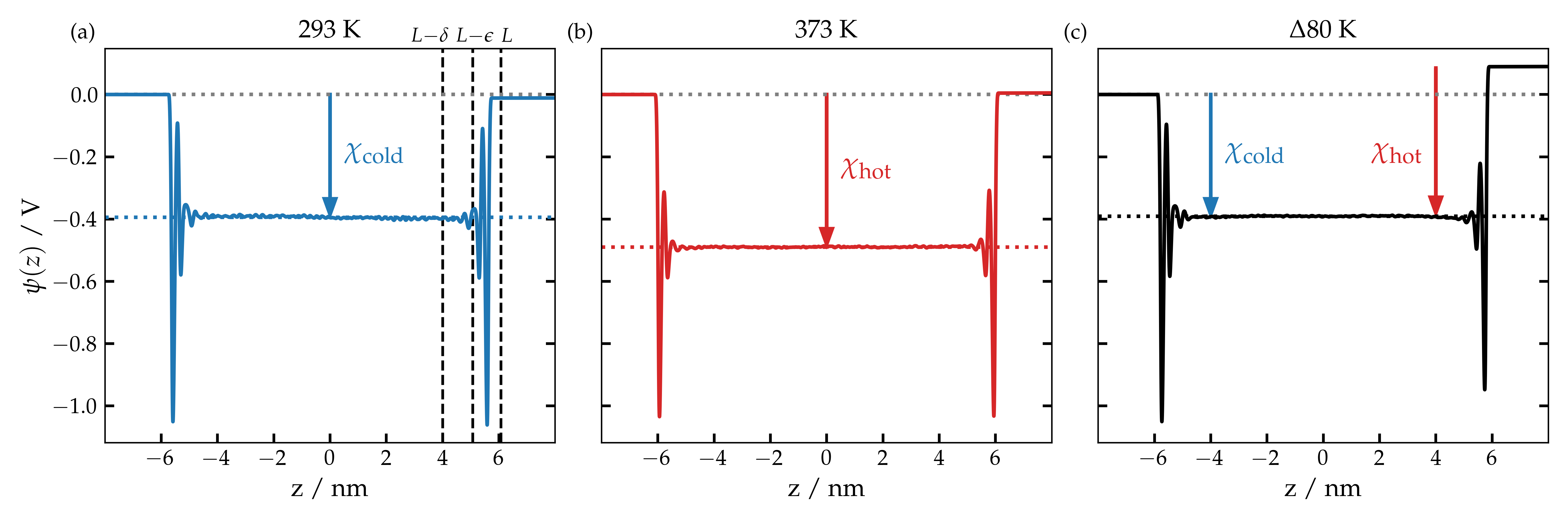}
    \caption{Potential profiles $\psi(z)$ of water at \SI{293}{\kelvin} (a), \SI{373}{\kelvin} (b), and with a temperature gradient, where the left side is at \SI{293}{\kelvin}, and the right side is at \SI{373}{\kelvin} (c).}
    \label{fig:U_profile_water}
\end{figure*}

Recent experiments on nonisothermal cells filled with aqueous \cite{sehnem_time-dependent_2021} and polymeric \cite{zhao_ionic_2016} electrolytes found Seebeck coefficients above $ \mathcal{S}=\SI{1}{\milli\volt\per\kelvin}$, much higher than predicted by \cref{eq:seebeck}.
Another experiment by \citet{xu_harvesting_2012} found potential differences between blocking electrodes held at different temperatures in separate electrolyte reservoirs, ruling out thermodiffusion as its cause in this case. 
Together these studies suggest that mechanisms beyond thermodiffusion contribute to the electrolyte Seebeck effect.

Molecular dynamics (MD) simulations may help identify these mechanisms.
One MD study determined $G^\mathrm{hyd}_i$ from simulations of isothermal systems at different temperatures \cite{rezende_franco_molecular_2021}.
Others simulated nonisothermal electrolytes \cite{romer_alkali_2013, di_lecce_computational_2017, rezende_franco_molecular_2021} and found that the thermal polarization of bulk water contributes to the Seebeck effect \cite{di_lecce_thermal_2018}.
None of these works accounted for electrodes---simulating open, periodic systems instead.
However, the electrolyte/electrode interface may well contribute to the Seebeck effect.
MD simulations revealed a potential drop $\chi$ of about \SI{-500}{\milli\volt} over water/vapor interfaces; $\chi$ varied with temperature by up to \SI{-1}{\milli\volt/\kelvin} \cite{chapman_polarisation_2022, kathmann_understanding_2011, sokhan_free_1997, harder_origin_2008}. 
If the surface potential at electrode/electrolyte interfaces also depends on temperature, holding two electrodes at different temperatures would produce different surface potentials and a potential difference between them.
Here, we study nonisothermal thermoelectric cells with polarizable electrodes to determine how the electrode/electrolyte interface contributes to the electrolyte Seebeck effect. 

We simulated pure water and aqueous CsF, KCl, NaCl, and LiI between two flat parallel blocking electrodes made from three graphene layers (see \cref{fig:picture_cell}).
All MD simulations were performed with the June 2022 version of the Large-scale Atomic/Molecular Massively Parallel Simulator (\textsc{lammps}) \cite{thompson_lammps_2022}.
We used the constant charge method of the \textsc{electrode} package \cite{Ahrens-Iwers2022} to keep each polarizable electrode overall uncharged while allowing the charge on them to fluctuate locally \cite{Tee2023}. 
We used different force fields for the SPC/E water \cite{berendsen_missing_1987}, graphene \cite{cheng_computer_1990}, and halide and alkali metal ions \cite{deublein_set_2012}.
Nonbonded interactions had a cutoff of \SI{1.2}{\nano\meter}, and the \textsc{shake} algorithm held bonds and angles of water molecules rigid \cite{ryckaert_numerical_1977}. 
A PPPM $k\mathrm{-space}$ solver with a relative accuracy of $10^{-6}$ computed the long-range interactions \cite{ahrens-iwers_constant_2021}. 
The system was periodic in the electrode plane, and a correction for slab geometries was used \cite{yeh_ewald_1999}. 
Initial simulations cells were around $\SI{2}{\nano\meter} \times \SI{2}{\nano\meter} \times \SI{16}{\nano\meter}$. 
A force on one of the electrodes equivalent to a pressure of \SI{1}{bar} adjusted the electrolyte's density. 
After letting the system equilibrate for a few nanoseconds, the distance $2L$ between the electrodes was between 11.4 to \SI{12.2}{\nano\meter}, and the position of the electrodes was fixed. 
Thermostats set the temperature of the electrolyte by global velocity rescaling with Hamiltonian dynamics \cite{bussi_canonical_2007} in two regions immediately next to the electrodes, both of width $\epsilon=\SI{1} {\nano\meter}$. 
Upon applying a temperature difference $\Delta T$, a linear temperature with a slope $\Delta T/[2(L-\epsilon)]$ developed between the thermostats in less than  \SI{1}{\nano\second}. 
All simulations were performed with a time step of \SI{1}{\femto\second} over \SI{30}{\nano\second}. 
For better statistics, simulations were repeated up to eight times with independent starting positions and flipped thermal gradients (thus up to 16 simulations).

We first discuss cells with 1440 water molecules and no ions.
We use a Cartesian coordinate system with $x$ and $y$ lying in the plane of the electrodes of surface area $A$; the coordinate $z$ runs from $-L$ to $L$ between the electrodes.
The water's partial charges cause a spatially varying charge density $\rho(z)$ and local potential $\psi(z)$, according to the twice-integrated Poisson equation \cite{di_lecce_thermal_2018, chapman_polarisation_2022},
\begin{equation}
    \psi(z)=-\frac{1}{\varepsilon_0} \int_{z_0}^z \int_{z_0}^{z^{\prime}} \rho \left(z^{\prime \prime}\right) \mathrm{d} z^{\prime \prime} \mathrm{d} z^{\prime},
\label{eq:potential-poission}
\end{equation}
where $\varepsilon_0$ is the vacuum permittivity, and $z_0$ is an arbitrary reference point left of the cell where we set $\psi(z_0)=0$.
We sample $\rho(z)$ from the MD simulations by time-averaging the partial charges in bins spanning the $xy$ plane and \SI{1}{\pico\meter} wide in the $z$-direction. 

\Cref{fig:U_profile_water} shows $\psi(z)$ for a ``cold" and ``hot" isothermal system at \SI{293}{\kelvin} [\cref{fig:U_profile_water}(a)] and \SI{373}{\kelvin} [\cref{fig:U_profile_water}(b)], respectively.
In both panels, $\psi(z)$ varies strongly near the electrodes but not in the cell's interior.
To characterize $\psi(z)$, we divide the cell into two interfacial regions $L-\delta<|z|<L$ of thickness $\delta$ and a bulk region where $|z|<L-\delta$.
We choose $\delta$ so that the bulk region is $|z|<\SI{4}{\nano\meter}$. 
From linear fits to the potential in the bulk $\psi_{\rm fit}(z)$ we determine the bulk potential drop, $\Delta\psi_\mathrm{bulk}=\psi_{\rm fit}(L-\delta)-\psi_{\rm fit}(\delta-L)$. 
Both \cref{fig:U_profile_water}(a) and (b) have $\Delta\psi_\mathrm{bulk}\approx\SI{0}{\milli\volt}$, so water molecules are not polarized in the bulk.
We define the surface potential drop $\chi$ at each interface as the difference between the electrode potential and the average potential at the edge of the bulk region, $\chi_{\pm}=\psi_{\rm fit}(\pm L\mp\delta)-\psi(\pm L)$. 
We observe similar surface potentials at both electrodes $|\chi_-|\approx |\chi_+|$, with $\chi_\mathrm{cold}= \SI{-391}{\milli\volt}$ and $\chi_\mathrm{hot}=\SI{-489}{\milli\volt}$. 
As a result, the potential difference $\Delta \psi = \psi(L) - \psi(-L)$ between electrodes, which can be partitioned as $\Delta \psi=\chi_{-}+\Delta\psi_\mathrm{bulk}-\chi_{+}$, is roughly zero.
In general, the temperature dependence of $\chi$ is non-linear.
\Cref{fig:surface_potential} in the Supplemental Material (SM) \cite{supplement_footnote} shows that $\chi$ decreases roughly linearly between 293 and \SI{373}{\kelvin} with $(\chi_\mathrm{hot} - \chi_\mathrm{cold}) / \SI{80}{\kelvin}=\SI{-1.22}{\milli\volt/\kelvin}$, slightly higher than reported values for the SPC/E water/vapor interface \cite{kathmann_understanding_2011, chapman_polarisation_2022}. 

\Cref{fig:U_profile_water}(c) shows $\psi(z)$ in a cell with the electrodes held at \SI{293}{\kelvin} and \SI{373}{\kelvin}, respectively.
The respective surface potentials are similar to those in panels \Cref{fig:U_profile_water}(a) and (b) and, therefore, depend on the \textit{local} temperature: $\chi_-\approx\chi_\mathrm{cold}$ and $\chi_+\approx\chi_\mathrm{hot}$.
As the two surface potentials no longer cancel, we observe a nonzero $\Delta \psi$ and a corresponding Seebeck coefficient $\mathcal{S}=\SI{-1.25}{\milli\volt/\kelvin}$ \footnote{For nonisothermal systems with electrodes at temperatures between \SI{293}{\kelvin} and \SI{373}{\kelvin}, we can estimate the surface potentials $\chi_\mathrm{cold}$ and $\chi_\mathrm{hot}$ and corresponding Seebeck coefficient $\approx\chi_\mathrm{cold}-\chi_\mathrm{hot}$ from an interpolation or fit to the $\chi$ of isothermal system plotted in \cref{fig:surface_potential} in the SM \cite{supplement_footnote}.}.
The nonisothermal cell also has a small bulk potential drop $\Delta\psi_\mathrm{bulk}\approx\SI{1}{\milli\volt}$, in line with previous MD studies \cite{di_lecce_thermal_2018, chapman_polarisation_2022, armstrong_temperature_2015} \footnote{Armstrong and Bresme \cite{armstrong_temperature_2015} found strong water polarization at high temperatures ($\sim \SI{550}{\kelvin}$). In the liquid range of water, the thermal polarization was much weaker, changing sign at \SI{320}{\kelvin}.}.
To separate surface and bulk contributions to $\mathcal{S}$, we introduce a corresponding surface Seebeck coefficient, $\mathcal{S}_\mathrm{surf}=(\chi_{+}-\chi_{-})/\Delta T$, and bulk Seebeck coefficient, $\mathcal{S}_\mathrm{bulk}= -\Delta\psi_\mathrm{bulk}/\Delta T_\mathrm{bulk}$, where $\Delta T_\mathrm{bulk}=\Delta T (L-\delta)/ (L-\epsilon)$ is the bulk temperature drop.
As $\Delta T_\mathrm{bulk}<\Delta T$, these definitions yield $\mathcal{S}\neq \mathcal{S}_\mathrm{bulk}+\mathcal{S}_\mathrm{surf}$.
\Cref{table:seebeck_data} lists all the resulting coefficients.

\begin{table*}
\begin{ruledtabular}
\centering
\caption{Potential drop and Seebeck coefficients of water and several electrolytes in different cell regions (electrodes held at \SI{293}{\kelvin} and \SI{373}{\kelvin}, respectively). 
Mean values and uncertainties are determined by block averaging 16 simulations. }
\begin{tabular}{lccccccccc} 
	&	$\Delta \sigma_\mathrm{LJ}$	&	$\chi_-$	&	$\Delta\psi_\mathrm{bulk}$	&	$\chi_+$	&	$\mathcal{S}$	&	$\mathcal{S}_\mathrm{surf}$	&	$\mathcal{S}_\mathrm{bulk}$	&	$\mathcal{S}_\mathrm{Agar}$	&	$\mathcal{S}_\mathrm{Marcus}$	\\
	&	{\AA}	&	\SI{}{\milli\volt}	&	\SI{}{\milli\volt}	&	\SI{}{\milli\volt}	&	\SI{}{\milli\volt/\kelvin}	&	\SI{}{\milli\volt/\kelvin}	&	\SI{}{\milli\volt/\kelvin}	&	\SI{}{\milli\volt/\kelvin}	&	\SI{}{\milli\volt/\kelvin}	\\
\hline
Water	&	n/a	&	$-395\pm3$	&	$1\pm4$	&	$-495\pm1$	&	$-1.25\pm0.05$	&	$-1.25\pm0.04$	&	$-0.01\pm0.05$	&	n/a\footnote{\citet{agar_single-ion_1989} reported $Q_i^\ast$ and \citet{marcus_simple_1994} $S^\mathrm{hyd}_i$ for OH$^{-}$ and H$_3$O$^{+}$, but we did not include these ions in our MD simulations.}	&	n/a$^\mathrm{a}$\\
\SI{1}{\molar} CsF	&	0.08	&	$-337\pm10$	&	$-31\pm13$	&	$-469\pm8$	&	$-1.26\pm0.12$	&	$-1.65\pm0.21$	&	$0.46\pm0.19$	&	$0.00$ &	$-0.40$	\\
\SI{1}{\molar} KCl	&	1.64	&	$-387\pm4$	&	$0\pm10$	&	$-490\pm3$	&	$-1.29\pm0.11$	&	$-1.29\pm0.06$	&	$0.00\pm0.15$	&	$0.03$	&	$-0.01$	\\
\SI{1}{\molar} NaCl	&	2.52	&	$-413\pm6$	&	$10\pm12$	&	$-511\pm5$	&	$-1.34\pm0.16$	&	$-1.22\pm0.09$	&	$-0.14\pm0.18$	&	$0.05$	&	$0.19$	\\
\SI{1}{\molar} LiI	&	2.90	&	$-412\pm4$	&	$14\pm10$	&	$-509\pm5$	&	$-1.39\pm0.13$	&	$-1.22\pm0.08$	&	$-0.20\pm0.14$	&	$0.03$	&	$0.55$	\\
\end{tabular}
\label{table:seebeck_data}
\end{ruledtabular}
\end{table*}

\begin{figure}
    \centering
    \includegraphics[width=\linewidth]{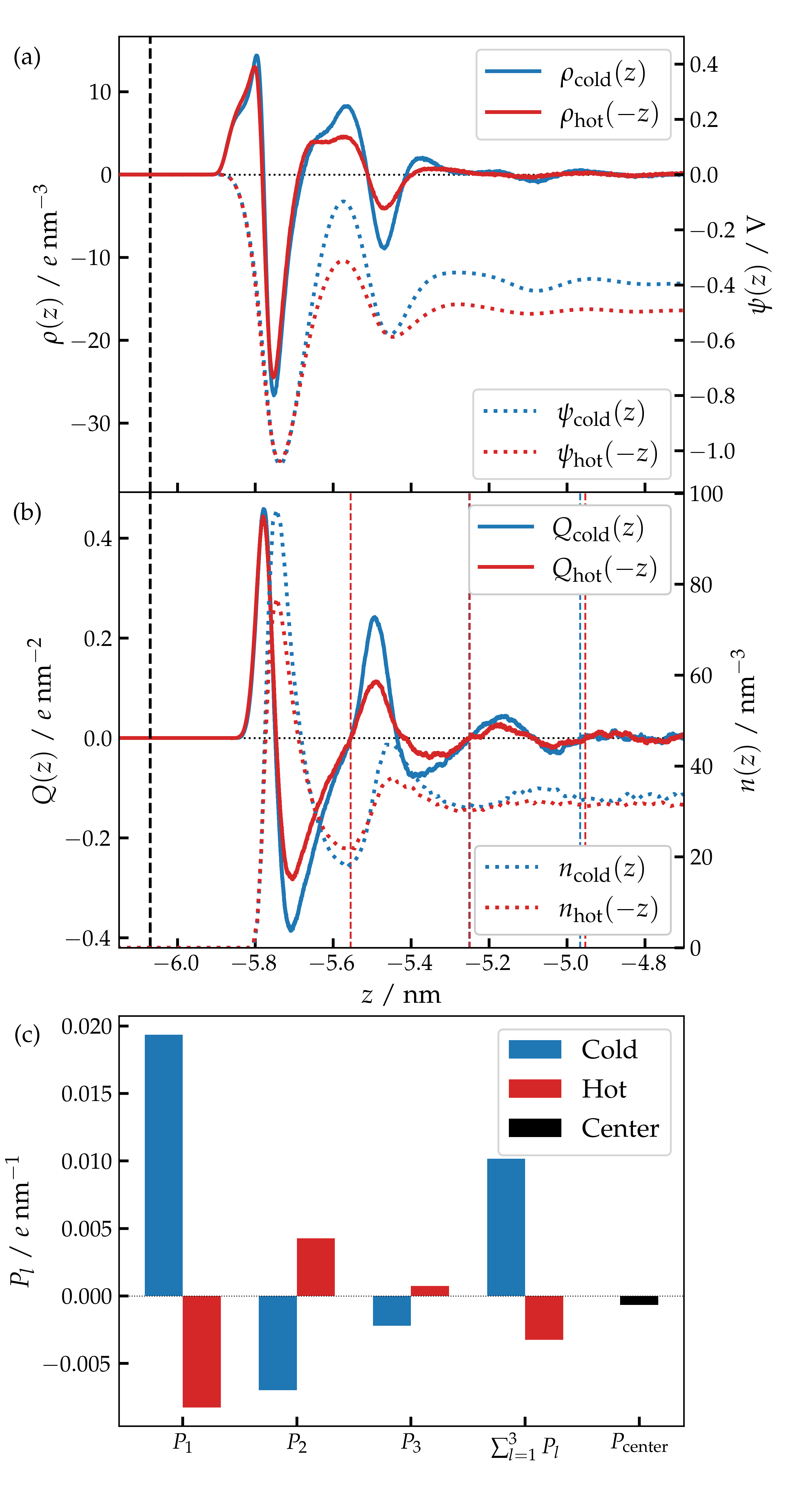}
    \caption{
    Further analysis of the water-filled cell of \cref{fig:U_profile_water}(c) with temperature varying from 293 to \SI{373}{\kelvin}. 
    (a) Charge density and local potential, where the hot side at \SI{373}{\kelvin} is mirrored around $z=0$, and its potential in vacuum shifted to \SI{0}{\volt}. The vertical black line indicates the position of the first graphene layer.
    (b) Cumulative charge (lines) and oxygen number density (dotted lines). 
    (c) Layer dipole moment of water in the first three neutral layers near the cold and hot sides of the cell, their sum, and the dipole moment of the bulk water in the remainder of the cell (black).}
    \label{fig:potential_contributions}
\end{figure}

We show in \cref{fig:potential_contributions}(a) the near-electrode potential profiles of \cref{fig:U_profile_water}(c) shifted to zero at the electrodes (dotted lines) and their underlying charge density profiles (solid lines).
The peaks and valleys of the charge density are inversely related to those of the mass density (see \cref{fig:density} in the SM \cite{supplement_footnote}): where oxygen dominates, the charge density is negative and the mass density high; where hydrogen dominates, the charge density is positive and the mass density low. 
\Cref{fig:potential_contributions} shows that the charge density $\rho(z)$ varies less at a higher temperature, changing the surface potential $\chi$.

To understand the temperature-dependent surface potential drop of water, we inserted multipole expansions of the charge density, $\rho(z)=\mathcal{M}(z)-\frac{\mathrm{d}}{\mathrm{d} z}\mathcal{P}_z(z)+\frac{\mathrm{d}^2 }{\mathrm{d}^2 z}\mathcal{Q}_{z z}(z)+\ldots$, 
into \cref{eq:potential-poission} to determine how monopolar ($\mathcal{M}$), dipolar ($\mathcal{P}_z$), and quadrupolar ($\mathcal{Q}_{z z}$) terms contribute to $\psi(z)$ \cite{Matsumoto1988, glosli_molecular_1996, sokhan_free_1997,Bonthuis2012, Cendagorta2015, armstrong_temperature_2015, Iriarte-Carretero2016, di_lecce_thermal_2018, Doyle2019, Gittus2020, chapman_polarisation_2022, becker_electrokinetic_2022}.
However, multipole expansions are not unique \cite{Wilson1989}: the terms $\mathcal{M}, \mathcal{P}_z$, and $\mathcal{Q}_{z z}$ contain an arbitrary reference point $z_m^\mathrm{ref}$ [see \cref{eq:multipole} in the SM \cite{supplement_footnote}]. 
\Cref{fig:psi_contributions} in the SM \cite{supplement_footnote} shows $\mathcal{P}_z$ and $\mathcal{Q}_{z z}$'s contributions to $\psi(z)$ for five different $z_m^\mathrm{ref}$ ($\mathcal{M}=0$ for water).
These data differ much so the different terms in a multipole expansion give little physical insight.
Nevertheless, $\mathcal{Q}_{z z}$'s contribution to the potential difference between two points is proportional to the difference in $\mathcal{Q}_{z z}(z)$ between those points \cite{Wilson1989}.
As $\mathcal{Q}_{z z}=0$ in the electrodes, the quadrupole density does not contribute to $\Delta \psi$.
Consequently, in water-filled cells, only $\mathcal{P}_z$ (and $\mathcal{M}$, if ions are added) contribute to $\Delta \psi$.

To unambiguously characterize the water's orientation, we consider charge-free layers that we define in terms of the cumulative charge
\begin{equation}\label{eq:cumulative_charge}
    Q(z) = \frac{1}{A} \int_{z_0}^z \mathrm{d}z^\prime \sum_{i=1}^{N_a} \delta(z^\prime - z_i) q_i\,,
\end{equation}
of $N_a$ atoms with partial charge $q_i$ and position $z_i$ along the $z$-axis.
\Cref{fig:potential_contributions}(b) shows $Q(z)$ for the nonisothermal cell of \cref{fig:U_profile_water}(c).
For this analysis, we moved the two hydrogen atoms of each water molecule to the point between them.
In this way, $Q(z)$ is a multiple of the oxygen charge and the points where $Q(z)=0$ are easy to identify.
For a single layer of water molecules, seen as a peak in the oxygen density, $Q(z)$ rises, drops and passes zero, and rises again.
Accordingly, we set the boundaries $\zeta_{l-1}$ and $\zeta_l$ of uncharged layers $l=1,2,3,\ldots$ at the second crossings of the cumulative charge, $Q(\zeta_{l-1})=Q(\zeta_l)=0$, and the boundary of the first layer at $\zeta_{l-1} = z_0$.
With this definition, the layer boundaries (vertical dotted lines) are close to minima in the oxygen number density (dotted lines).
\Cref{fig:potential_contributions}(b) clearly shows three layers near both electrodes; the boundaries of further layers are difficult to determine due to a weak signal-to-noise ratio.
Accordingly, we only consider the first three layers near both electrodes and combine the rest of the cell in one big central layer.
The distribution of atomic charges in layer $l$ is captured by its layer dipole moment,
\begin{equation}\label{eq:layer-dipole}
    P_{l} = \frac{1}{A} \int_{\zeta_{l-1}}^{\zeta_{l}} \mathrm{d}z \sum_{i=1}^{N_a} \delta(z-z_i) z_i q_i\,.
\end{equation}
Note that neither \cref{eq:cumulative_charge,eq:layer-dipole} contains any arbitrary reference points.
\Cref{fig:potential_contributions}(c) shows $P_1, P_2$, and $P_3$ on the cold (blue) and hot (red) sides of the cell of \cref{fig:U_profile_water}(c).
$P_l$ takes positive and negative values, and has larger absolute values at the cold side. 
The figure also shows the sum of the dipoles, three times larger near the cold side, and the dipole moment $P_{\rm center}$ of the rest of the cell, which is much smaller than $P_1, P_2$, and $P_3$.
Summing all $P_l$ yields a nonzero overall dipole moment and associated potential drop $\Delta \psi=\sum_l P_l/\varepsilon_0$.
Hence, \cref{fig:potential_contributions}(c)  shows that an increase in temperature leads to decreased water ordering and, in turn, a net dipole and potential difference.

We now turn to electrolytes with 1332 water molecules and 24 ion pairs of CsF, KCl, NaCl, or LiI between electrodes held at \SI{293}{\kelvin} and \SI{373}{\kelvin}. 
The ion's sizes are set by the $\sigma^{\pm}_\mathrm{LJ}$ parameter of the Lennard-Jones force field; \cref{table:seebeck_data} lists the differences in ion sizes $\Delta\sigma_\mathrm{LJ}=\sigma^{-}_\mathrm{LJ}-\sigma^{+}_\mathrm{LJ}$ for each electrolyte. 
The table also presents MD results for the potential drops $\chi_-$, $\Delta\psi_\mathrm{bulk}$, and $\chi_+$, which we use to determine the Seebeck coefficients $\mathcal{S}, \mathcal{S}_\mathrm{surf}$, and $\mathcal{S}_\mathrm{bulk}$.
Last, \cref{table:seebeck_data} contains predictions $\mathcal{S}_\mathrm{Agar}$ and $\mathcal{S}_\mathrm{Marcus}$, for which we inserted $Q_i^\ast$ data from Table 1 of \citet{agar_single-ion_1989} and $S^\mathrm{hyd}_i$ data from Table 2 of \citet{marcus_simple_1994} into \cref{eq:soret-entropy,eq:seebeck}.

\Cref{table:seebeck_data} shows us the following. 
First, our MD simulations yield $\mathcal{S}$ of the same size but opposite sign as experiments \cite{sehnem_time-dependent_2021}.
This discrepancy may be caused by the different water structure near different electrodes; the experiments of Ref.~\cite{sehnem_time-dependent_2021} employed titanium, whereas our simulations concerned graphite electrodes.
Second, \Cref{table:seebeck_data} shows that $|\mathcal{S}|\gg |\mathcal{S}_\mathrm{bulk}|$ for all electrolytes, $|\mathcal{S}|\approx |\mathcal{S}_\mathrm{surf}|$ for most electrolytes, and that adding ions to water leads to a nonsignificant decrease in $\mathcal{S}$ for all electrolytes.
These observations suggest that the electrolyte Seebeck effect is more caused by interfacial electrolyte ordering than thermodiffusion. 
Third, for all ion pairs except CsF, $\mathcal{S}_\mathrm{surf}$ values lie around $\mathcal{S}=\SI{-1.25}{\milli\volt/\kelvin}$ of pure water.
This does not mean, however, that the electrolyte Seebeck effect is caused only by interfacial water structure.
Adding ions to water changes $\chi_-$ and $\chi_+$, but usually by the same amount, so the shifts in the surface potential drops cancel in $\mathcal{S}_\mathrm{surf}=(\chi_{-}-\chi_{+})/\Delta T$.
\Cref{fig:ions_density} in the SM \cite{supplement_footnote} shows that the CsF anomaly is caused by a strong Cs\textsuperscript{+} localization in the second water layer near the cold electrode, not present near the hot electrode.
Fourth, $\mathcal{S}_\mathrm{Agar}$ and $\mathcal{S}_\mathrm{Marcus}$ correspond to bulk ionic thermodiffusion, but neither agrees with $\mathcal{S}_\mathrm{bulk}$.
Marcus's theory predicts different sized ions to have different $S^\mathrm{hyd}_i$, so thermodiffusion should contribute strongly to the Seebeck coefficient of electrolytes with small and large ions \cite{marcus_simple_1994}.
We indeed see that $\mathcal{S}_\mathrm{bulk}$ decreases monotonously with $\Delta \sigma_\mathrm{LJ}$; Marcus's theory predicts a monotonous increase instead.
The discrepancies between $\mathcal{S}_\mathrm{bulk}$ and $\mathcal{S}_\mathrm{Agar}$ and $\mathcal{S}_\mathrm{Marcus}$ 
may be caused by temperature and concentration dependence of $\mathcal{S}_\mathrm{bulk}$.
Previous MD simulations found $\mathcal{S}_\mathrm{bulk}$ to depend strongly on temperature; for LiCl, $\mathcal{S}_\mathrm{bulk}$ changed sign around \SI{305}{\kelvin} \cite{di_lecce_thermal_2018}.
Moreover, Agar's and Marcus's data concerned electrolytes at infinite dilution, while our simulations discussed so far were at \SI{1}{\molar}.
In the SM \cite{supplement_footnote} (see \cref{fig:S_bulk_concentration}), we consider different salt concentrations and find that $\mathcal{S}_\mathrm{bulk}$ generally increases with concentration.
However, even at the smallest concentrations studied there (0.5\,M),  discrepancies between $\mathcal{S}_\mathrm{bulk}$ and $\mathcal{S}_\mathrm{Agar}$ and $\mathcal{S}_\mathrm{Marcus}$ persisted.
Finally, the small values of $\mathcal{S}_\mathrm{Agar}$ and $\mathcal{S}_\mathrm{Marcus}$ further support our conclusion that the large $\mathcal{S}$ in our simulations cannot be explained by ionic thermodiffusion alone.
%

In summary, we presented the first MD simulations of nonisothermal electrolyte-filled cells with explicit electrodes.
In our MD simulations, (i) the Seebeck effect of aqueous electrolytes is caused more by interfacial electrolyte structure than by ionic thermodiffusion;
(ii) adding ions to pure water changes the interfacial electrolyte structure \cite{Hedley2023}, but usually to the same degree near the cold and hot electrodes, leaving $\mathcal{S}_\mathrm{surf}$ unaffected;
(iii) current models do not capture the thermodiffusion of dense electrolytes.
Hence, further work is needed to fully understand the electrolyte Seebeck effect.

So far, many experiments concerned polyelectrolytes, ionic polymer gels, and ionic liquids \cite{zhao2021ionic}.
Experiments, theory, and simulations are more likely to meet for systems with simpler components, for instance, aqueous electrolytes, only a few of which have been studied \cite{sehnem_time-dependent_2021}.
Next, it would be interesting to go beyond our idealized flat-electrodes setup.
An electrode's shape, surface defects, and functional groups should affect the nearby electrolyte structure.
Tailoring these properties could boost a system's Seebeck coefficient, making it more attractive for applications.

\bibliography{bib.bib}

\clearpage
\setcounter{equation}{0}
\setcounter{figure}{0}
\setcounter{section}{0} 
\renewcommand{\thefigure}{{S}\arabic{figure}}
\renewcommand{\thesection}{{S}\arabic{section}}
\renewcommand{\theequation}{{S}\arabic{equation}}
\newpage
\onecolumngrid
\begin{center}
{\bf SUPPLEMENTARY MATERIAL to: Water, not salt, causes most of the Seebeck effect of nonisothermal aqueous electrolytes}\\\vspace{0.2cm}
Ole Nickel, Ludwig J. V. Ahrens-Iwers, Robert H. Mei{\ss}ner, Mathijs Janssen
\date{\today}
\end{center}

\section{Surface potentials of water}

\begin{figure}[H]
\centering
\includegraphics[width=0.6\linewidth]{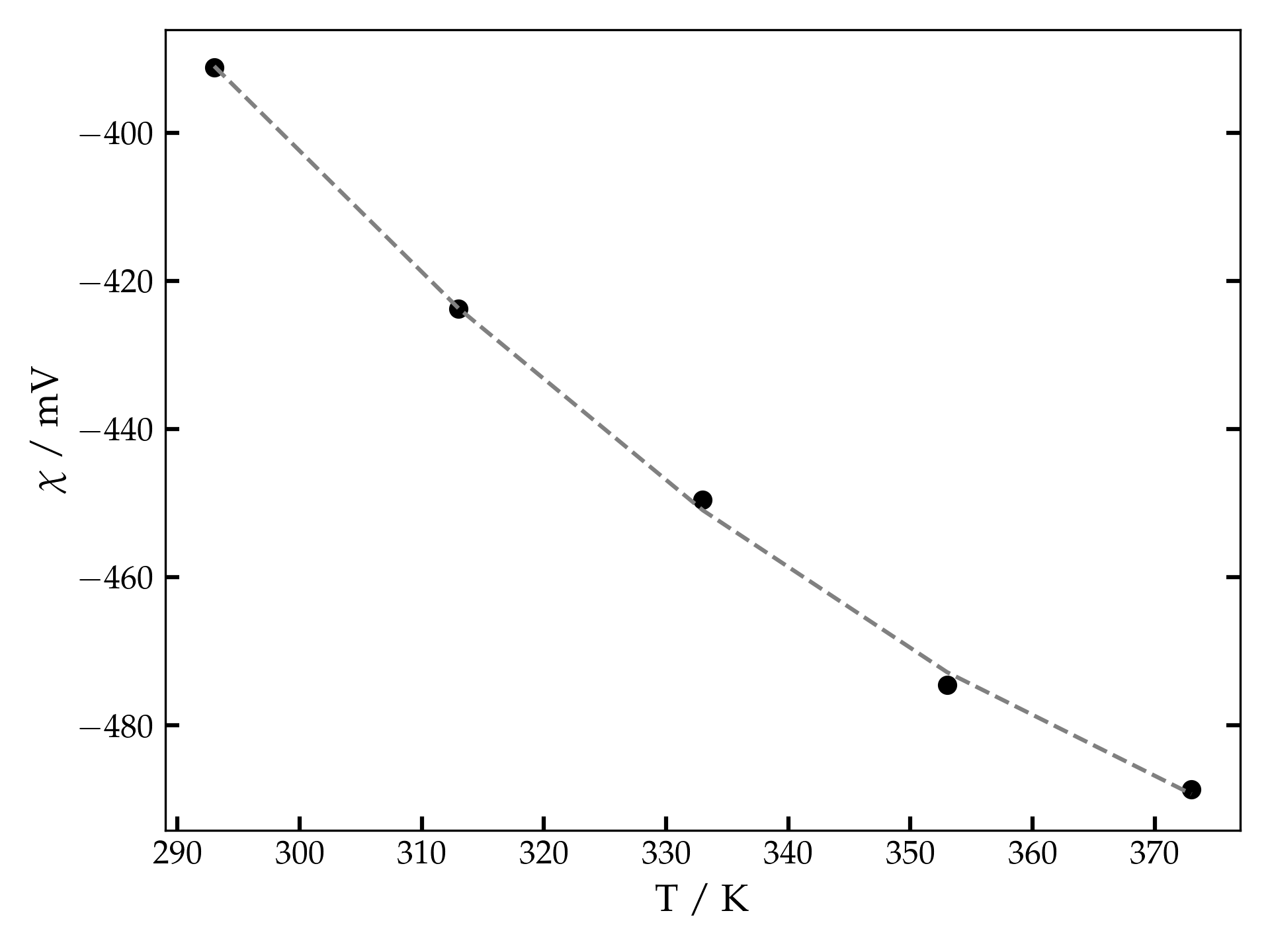}
\caption{The surface potential of the SPC/E water/graphene interface at different temperatures.}
\label{fig:surface_potential}
\end{figure}

\Cref{fig:surface_potential} shows the surface potential $\chi$ for SPC/E water for several isothermal systems between \SI{293}{\kelvin} to \SI{373}{\kelvin}. 
Similar to \textcite{chapman_polarisation_2022,sokhan_free_1997}, where the surface potential is discussed for different water models at the water/vapor interface, we find that $\chi$
decreases monotonously and the slope is reduced at higher temperatures.
Hence, the Seebeck coefficient becomes larger when only smaller temperature differences at low temperatures are used.
We fitted the data in \cref{fig:surface_potential} by a quadratic function,
\begin{equation}
\chi = a + b\cdot(T-c)^2,
\label{eq:quadratic_fit}
\end{equation}
with the fit parameters $a=\SI{-507}{\milli\volt}$, $b=\SI{0.0068}{\milli\volt}$, and $c=\SI{424}{\kelvin}$.
Doing the parabolic fit in this way shows that $\chi$ opens upward ($b>0$) and reaches a minimum of \SI{-507}{\milli\volt} at \SI{424}{\kelvin}, which is close to prior MD simulation results \cite{chapman_polarisation_2022, sokhan_free_1997}.

\section{Mass density profile}

\Cref{{fig:density}} shows the mass density profile of water held between electrodes at \SI{293}{\kelvin} (left) and \SI{373}{\kelvin} (right).
Density peaks near both electrodes are more pronounced near the cold than the hot side.
Furthermore, the bulk density decreases from \SI{0.993}{\gram\per\centi\cubic\metre} to \SI{0.950}{\gram\per\centi\cubic\metre} in proportion to the local temperature, demonstrating the effectiveness of our density equilibration by applying a pressure of \SI{1}{\bar} to one of the electrodes to adjust the density.
\begin{figure}[H]
\centering
\includegraphics[width=1\linewidth]{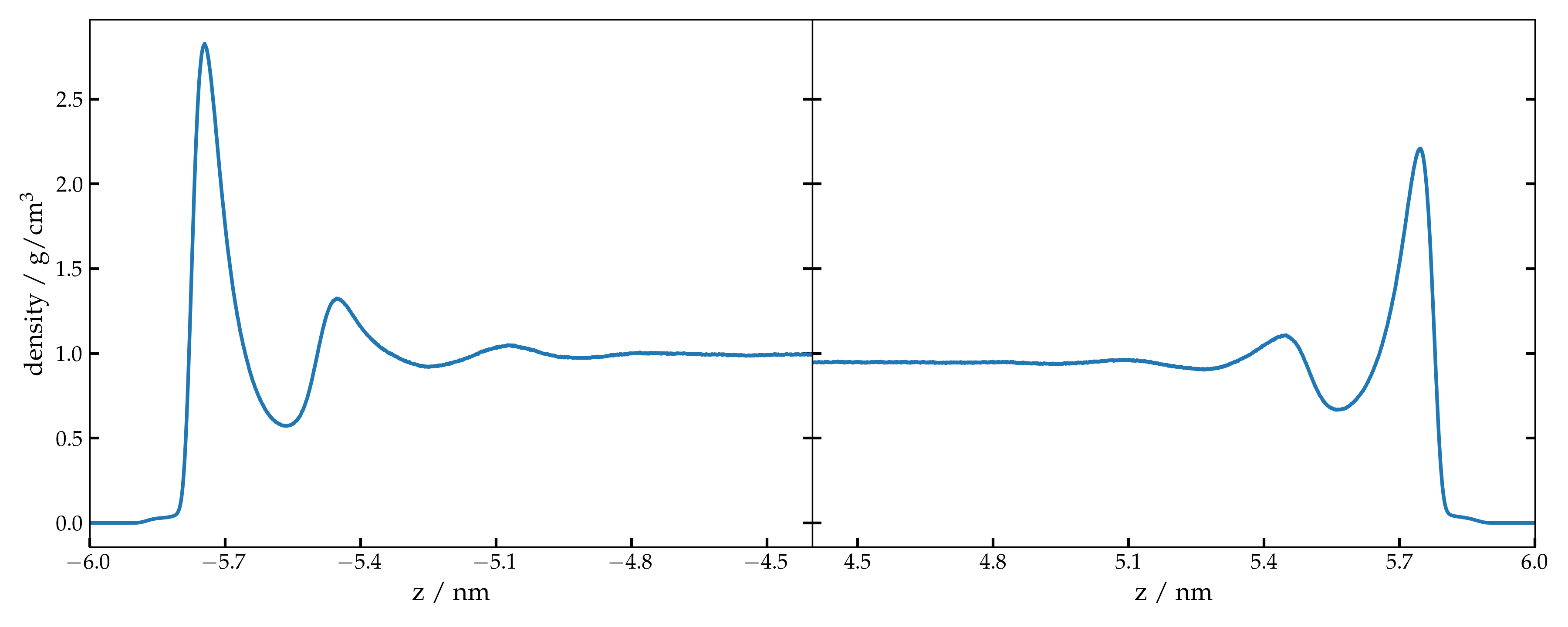}
\caption{Mass density profile of water, where the left side is held at \SI{293}{\kelvin} and the right side is held at \SI{373}{\kelvin}.}
\label{fig:density}
\end{figure}
\section{Multipole contributions to potential}

To analyze the charge density $\rho(z)$, one can perform a multipole expansion \cite{glosli_molecular_1996, di_lecce_thermal_2018}:
\begin{subequations}\label{eq:multipole}
\begin{align}
    \rho(z)&=\mathcal{M}(z)-\frac{\mathrm{d}}{\mathrm{d} z}\mathcal{P}_z(z)+\frac{\mathrm{d}^2 \mathcal{Q}_{z z}(z)}{\mathrm{d}^2 z}+\ldots,          \label{eq:multipole_expansion}\\
    \mathcal{M}(z) &= \frac{1}{A}\left\langle\sum_{m=1}^{N} \delta\left(z-z_m^\mathrm{ref}\right) \sum_{j=1}^{N_m} q_{mj}\right\rangle \label{eq:monopol} \\
    \mathcal{P}_z(z) &= \frac{1}{A}\left\langle\sum_{m=1}^{N} \delta\left(z-z_m^\mathrm{ref}\right) \sum_{j=1}^{N_m} q_{mj} z_{mj}\right\rangle \label{eq:dipol}\\
    \mathcal{Q}_{z z}(z) &= \frac{1}{A}\left\langle\sum_{m=1}^{N} \delta\left(z-z_m^\mathrm{ref}\right) \sum_{j=1}^{N_m} \frac{1}{2} q_{mj} z_{mj}^2\right\rangle,\label{eq:quadrupol}
\end{align}
\end{subequations}
where $\mathcal{M}(z)$, $\mathcal{P}_z(z)$, and $\mathcal{Q}_{z z}(z)$ are monopolar, dipolar and quadrupolar terms. 
Moreover, $A$ is the electrode area, and $N$ is the number of molecules (neutral and ionic), where the index $m$ labels each molecule.
Molecule $m$ has $N_m$ atoms, where the subscript $j$ labels all atoms within the molecule.
$z_m^\mathrm{ref}$ is a reference point in molecule $m$, which can be freely chosen, for instance, the oxygen atom in a water molecule. 
$z_{mj}$ is the $z$ coordinate of atom $j$ in site $m$ in the molecular frame of reference and $q_{mj}$ is the charge of atom $j$ in molecule $m$.
Brackets denote time averages over the simulation.
 
\begin{figure}
\centering
\includegraphics[width=\linewidth]{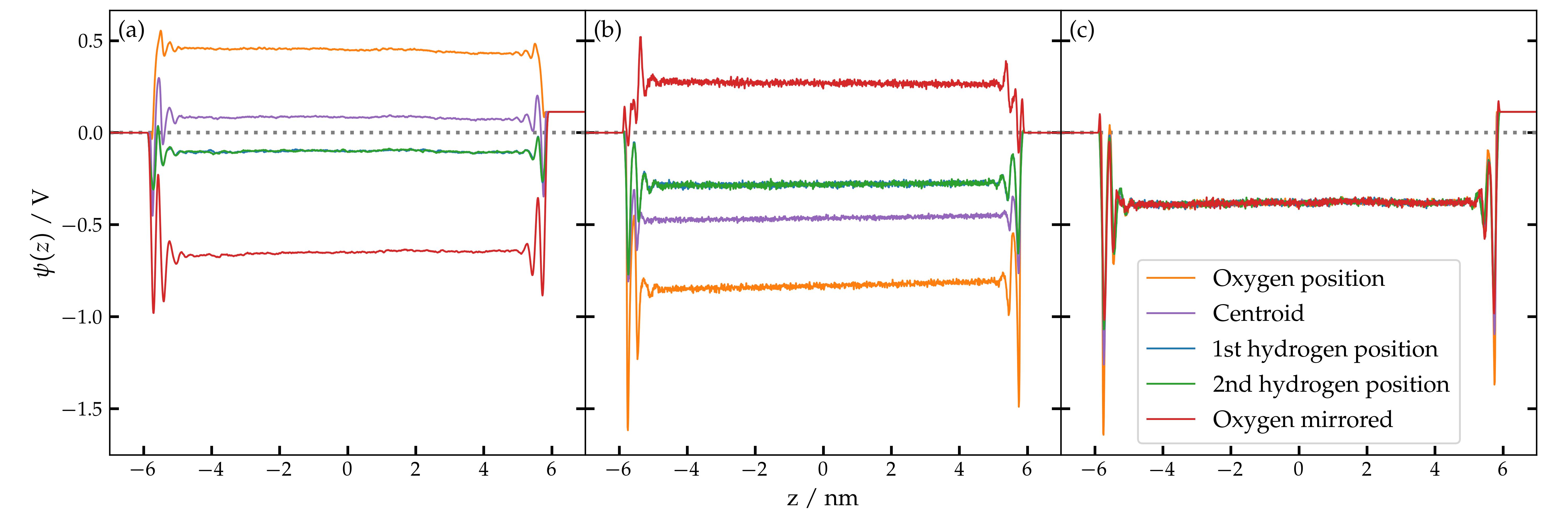}
\caption{
    Potential profiles for the same system as in \cref{fig:U_profile_water}c in the main text.
    Contributions to the potential from (a) the dipole moment, (b) the quadrupole moment, and (c) their sum. The contributions are shown for five choices of the reference point $z_m^\mathrm{ref}$: the position of each atom $z_m^\mathrm{ref} = z^\prime_{m j}$ with $j \in \{0,1,2\}$ and with the absolute atom positions $z^\prime_{mj} = z_m^\mathrm{ref} + z_{mj}$, the centroid of the three atom positions $z_m^\mathrm{ref} = \sum_j z^\prime_{mj}/3$ and the oxygen mirrored across the hydrogen atoms $z_m^\mathrm{ref} = -z^\prime_{m, \text{oxygen}} + \sum_{j \in \{1,2\}} z^\prime_{m j}$.
    These data correspond to a \SI{10}{\nano\second} simulation, where the position of every atom is sampled every \SI{1}{\pico\second} to calculate $\mathcal{P}_z(z)$ and $\mathcal{Q}_{z z}(z)$.}
\label{fig:psi_contributions}
\end{figure}

Studies of atomistic models often show local dipole and quadrupole contributions,
\begin{subequations}\label{eq:contributions}
\begin{align}
    \psi_\mathcal{P}(z) &= \frac{1}{\varepsilon_0} \int_{z_0}^z \mathrm d z^\prime \mathcal{P}_z(z^\prime) \label{eq:dipole_contribution},\\
    \psi_\mathcal{Q}(z) &= - \frac{1}{\varepsilon_0} \left[ \mathcal{Q}_{zz}(z) - \mathcal{Q}_{zz}(z_0) \right],\label{eq:quadrupole_contribution}
\end{align}
\end{subequations}
to the electrostatic potential \cite{Matsumoto1988, glosli_molecular_1996, sokhan_free_1997,Bonthuis2012, Cendagorta2015, Iriarte-Carretero2016, di_lecce_thermal_2018, Doyle2019, Gittus2020,  chapman_polarisation_2022, becker_electrokinetic_2022,armstrong_temperature_2015}. 
However, \citet{Wilson1989} pointed out that ``the integral of the molecular dipole density $\mathcal{P}_z (z)$ (and, thus, the electrostatic potential [...]) generally does depend on the choice of center at which the molecular dipole is located, even for the neutral molecule systems''.
We demonstrate the validity of that statement for water with a temperature difference from \SI{293}{\kelvin} to \SI{373}{\kelvin}. 
\Cref{fig:psi_contributions} shows  $\psi_\mathcal{P}(z)$ and $ \psi_\mathcal{Q}(z)$ for five different $z_m^\mathrm{ref}$: the three atomic position in a water molecule, the centroid of atom positions, and the oxygen position mirrored across the hydrogen atoms.
While  $\psi_\mathcal{P}(z)$ and $ \psi_\mathcal{Q}(z)$ change drastically between the different $z_m^\mathrm{ref}$ choices, their sum is hardly affected (we attribute the differences between the sums of contributions in \cref{fig:psi_contributions}(c) to numerical issues).
For unusual $z_m^\mathrm{ref}$ choices such as the hydrogen positions or the oxygen position mirrored across the hydrogen atoms, $\psi_\mathcal{P}(z)$ has an opposite sign compared to the more common choices of the oxygen position or the centroid.
The reason for these changes is that shifting the reference in the direction of the dipole will move molecules across the integration limit $z$ in \cref{eq:dipole_contribution} depending on the value of their dipole.
In fact, one can show that for shift $\Delta z_m^\mathrm{ref}$ in the reference, a term proportional to $\Delta z_m^\mathrm{ref} \sum_j q_{mj} z_{mj}$ is added to \cref{eq:dipole_contribution} and subtracted from \cref{eq:quadrupole_contribution}.
As there is no obvious choice for the reference point,  $\psi_\mathcal{P}(z)$ and $\psi_\mathcal{Q}(z)$ carry little informative value.

\section{Ionic charge density in nonisothermal cells}

\begin{figure}[H]
\centering
\includegraphics[width=\linewidth]{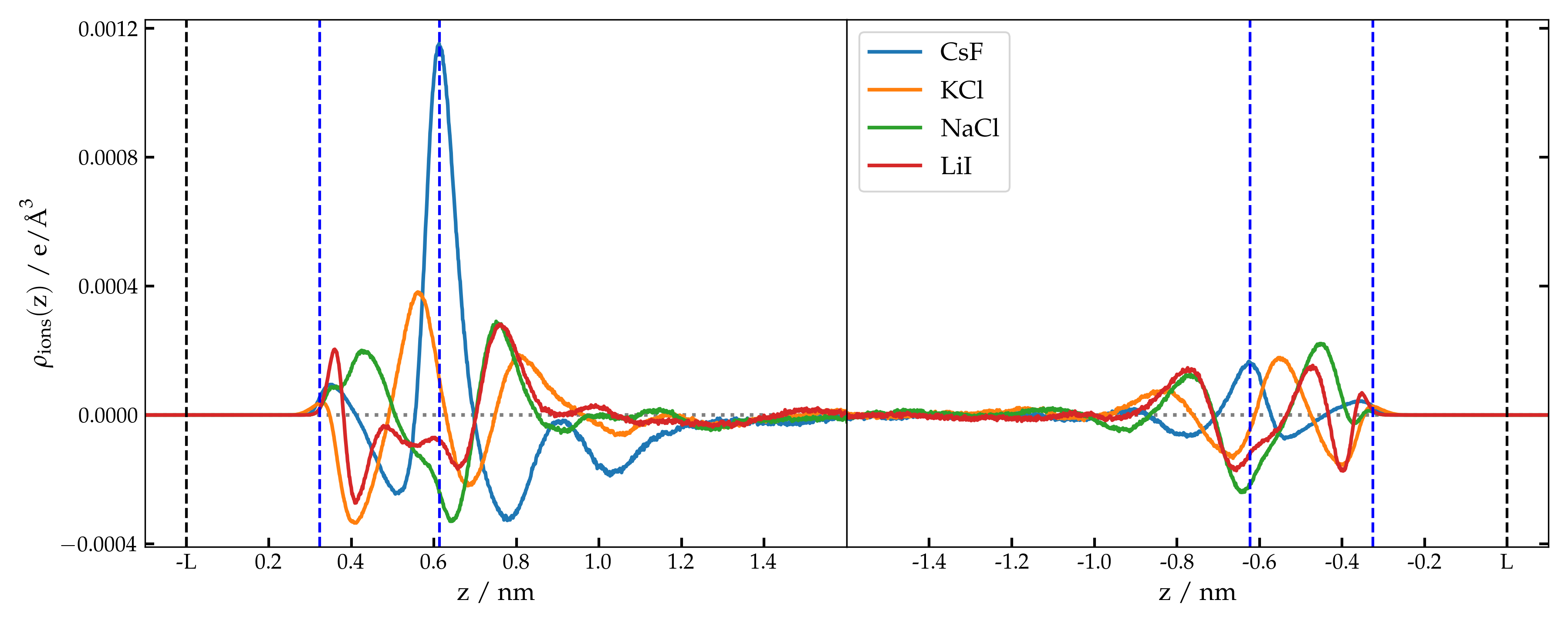}
\caption{Ionic charge density  $\rho_\mathrm{ions}(z)$ of different electrolytes held between an electrodes at 293 and \SI{373}{\kelvin} at left and right side. The electrodes at $-L$ and $L$ are in indicated by vertical black dashed lines; vertical blue dashed lines indicate the positions of the mass density peaks of water.}
\label{fig:ions_density}
\end{figure}

\Cref{fig:ions_density} shows the charge densities of just the ions  $\rho_\mathrm{ions}(z)$ neglecting the water in a nonisothermal cell.
All electrolytes have charge density maxima at different positions.
The location of these maxima depends on the cation and anion's preferred location in the water layers, which is mostly based on the size of the ion and its hydration shell.
Maxima signal that the cation is preferentially located there and minima mean that the anion is located there.
Near the cold electrode, KCl, NaCl, and LiI are less layered than CsF.
This explains their relatively small effect on the surface potential.
CsF layers strongly at the cold side, with caesium ions located in the second water layer at the water's mass density peak (blue vertical line).
The strong charge oscillations of CsF near the cold electrode cause a large increase in $\chi_-$ compared to the surface pure water.
CsF is much less layered near the hot interface, explaining the difference between $\chi_-$ and $\chi_+$ for this electrolyte.
From the singular ions densities, we have seen that both cations and anions move towards the cold electrode, leading to a pileup in its vicinity as expected.

\section{Influence of ion concentration on the Seebeck coefficient}

\begin{figure}[H]
\centering
\includegraphics[width=0.6\linewidth]{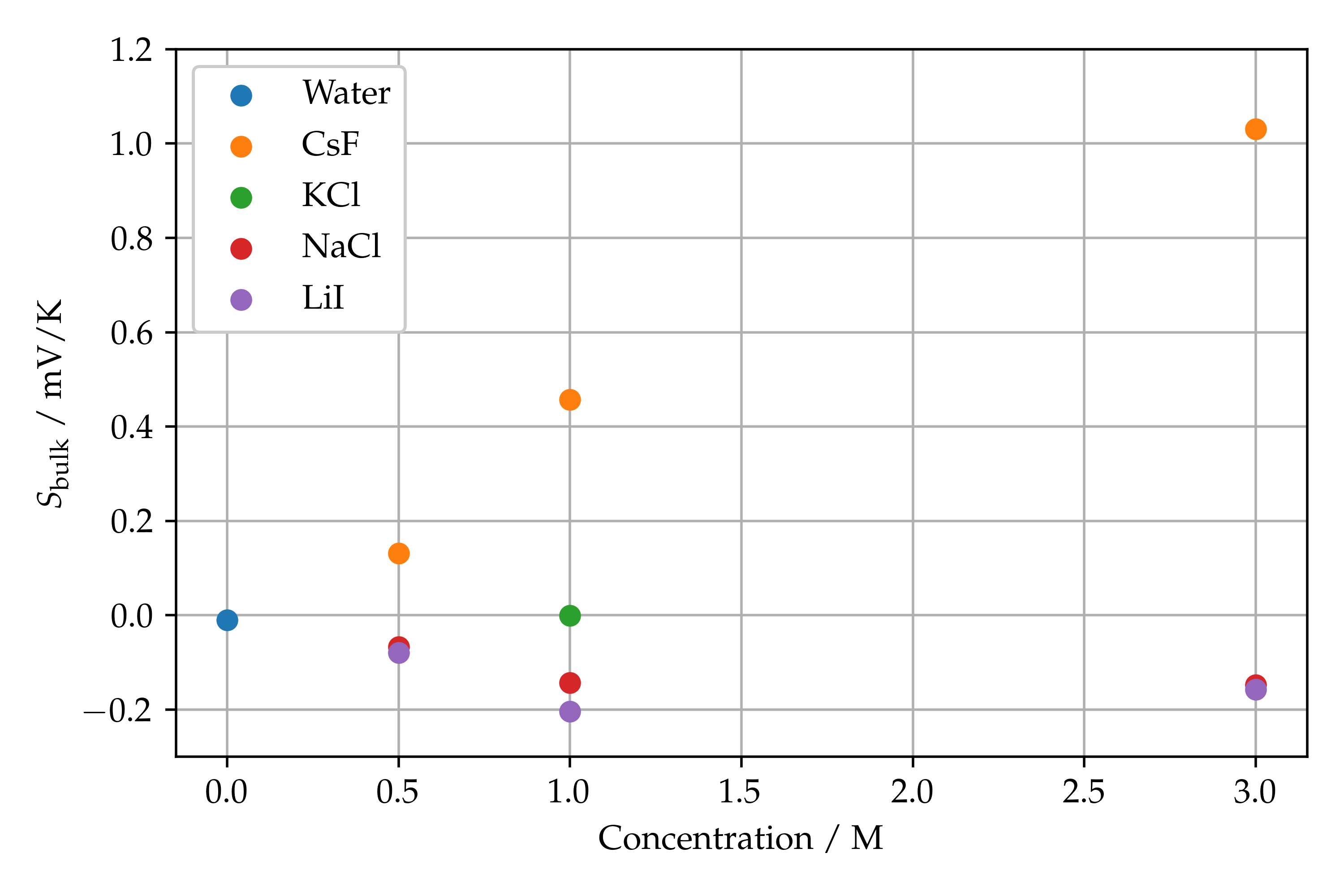}
\caption{Plot of $\mathcal{S}_\mathrm{bulk}$ for different electrolytes at different concentrations.}
\label{fig:S_bulk_concentration}
\end{figure}

We simulate further electrolytes with lower and higher concentrations and show the results in \cref{table:seebeck_concentration}.
We see changes in the surface potentials $\chi_{+}$ and $\chi_{-}$. 
NaCl and LiI values increase with concentration, so the overall effect on the $\mathcal{S}_\mathrm{surf}$ is still small.
For CsF, $\mathcal{S}_\mathrm{surf}$ increases significantly with concentration. 
This is again due to the strong localization of Cs$^+$ in the second water layer on the cold electrode.
We also find that $\mathcal{S}_\mathrm{bulk}$ depends on concentration (see \cref{fig:S_bulk_concentration}), and its absolute value generally increases with concentration.
This agrees with \citet{di_lecce_computational_2017}, who found a similar trend for LiCl at \SI{240}{\kelvin}.

\begin{table}[H]
\begin{ruledtabular}
\centering
\caption{Potential drop and Seebeck coefficients for electrolytes at different concentrations (electrodes held at \SI{293}{\kelvin} and \SI{373}{\kelvin}, respectively). Mean values and uncertainties are determined by block averaging 8 to 16 simulations.}
\begin{tabular}{lccccccc} 
	&	$\Delta \sigma_\mathrm{LJ}$	&	$\chi_-$	&	$\Delta\psi_\mathrm{bulk}$	&	$\chi_+$	&	$S$	&	$S_\mathrm{surf}$	&	$S_\mathrm{bulk}$	\\
	&	{\AA}	&	\SI{}{\milli\volt}	&	\SI{}{\milli\volt}	&	\SI{}{\milli\volt}	&	\SI{}{\milli\volt/\kelvin}	&	\SI{}{\milli\volt/\kelvin}	&	\SI{}{\milli\volt/\kelvin}	\\
\hline															
\SI{0.5}{\molar} CsF	&	0.08	&	$-372\pm3$	&	$-10\pm11$	&	$-481\pm4$	&	$-1.23\pm0.09$	&	$-1.36\pm0.08$	&	$0.13\pm0.17$	\\
\SI{0.5}{\molar} NaCl	&	2.52	&	$-411\pm3$	&	$5\pm13$	&	$-504\pm4$	&	$-1.23\pm0.19$	&	$-1.17\pm0.07$	&	$-0.07\pm0.19$	\\
\SI{0.5}{\molar} LiI	&	2.9	&	$-407\pm4$	&	$6\pm13$	&	$-502\pm2$	&	$-1.27\pm0.18$	&	$-1.20\pm0.04$	&	$-0.08\pm0.19$	\\	
\SI{3}{\molar} CsF	&	0.08	&	$-218\pm12$	&	$-75\pm23$	&	$-398\pm10$	&	$-1.31\pm0.07$	&	$-2.25\pm0.27$	&	$1.03\pm0.35$	\\
\SI{3}{\molar} NaCl	&	2.52	&	$-445\pm4$	&	$11\pm7$	&	$-540\pm4$	&	$-1.32\pm0.07$	&	$-1.18\pm0.09$	&	$-0.15\pm0.10$	\\
\SI{3}{\molar} LiI	&	2.9	&	$-437\pm4$	&	$11\pm6$	&	$-534\pm4$	&	$-1.35\pm0.08$	&	$-1.22\pm0.09$	&	$-0.16\pm0.08$	\\
\end{tabular}
\label{table:seebeck_concentration}
\end{ruledtabular}
\end{table}

\end{document}